\documentclass[a4paper,12pt]{article}

\usepackage[centertags]{amsmath}
\usepackage{amsfonts}
\usepackage{amssymb}
\usepackage{amsthm}
\usepackage{newlfont}
\usepackage[english]{babel}
\usepackage{enumerate}

\newcommand{\Lam}{\Lambda}

\newcommand{\eps}{\varepsilon}

\newcommand{\strl}{ {|}}
\newcommand{\projx}{ {x^n}}

\theoremstyle{plain}
  \newtheorem{theorem}{Theorem}[section]

  \newtheorem{lemma}[theorem]{Lemma}
\theoremstyle{definition}
  
  \newtheorem{assumption}[theorem]{Assumption}
\theoremstyle{remark}

\numberwithin{equation}{section}

 \DeclareMathOperator{\supp}{Supp}

\let\al=\alpha \let\be=\beta  \let\ep=\epsilon
   
  \let\om=\omega 
\let\si=\sigma

  \let\La=\Lambda \let\Om=\Omega

\newcommand{\caI}{{\mathcal I}}

\newcommand{\caM}{{\mathcal M}}

\newcommand{\bbN}{{\mathbb N}}

\newcommand{\bbR}{{\mathbb R}}
\newcommand{\bbS}{{\mathbb S}}

\newcommand{\opunit}{\text{1}\kern-0.22em\text{l}}

\newcommand{\e}{{\mathrm e}}
\renewcommand{\i}{{\mathrm i}}
\renewcommand{\d}{{\mathrm d}}

\newcommand{\beq}{ \begin{equation} }
\newcommand{\eeq}{ \end{equation} }
\newcommand{\bet}{ \begin{theorem} }
\newcommand{\eet}{ \end{theorem} }

\newcommand{\projectedgoodrays}{\widehat{\Om}_0''}
\newcommand{\projectedmarginalrays}{\widehat{\Om}_0'}

 \newcounter{smallarabics}
\newenvironment{arabicenumerate}
{\begin{list}{{\normalfont\textrm{\arabic{smallarabics})}}}
  {\usecounter{smallarabics}\setlength{\itemindent}{0cm}
  \setlength{\leftmargin}{5ex}\setlength{\labelwidth}{4ex}
  \setlength{\topsep}{0.75\parsep}\setlength{\partopsep}{0ex}
   \setlength{\itemsep}{0ex}}}
{\end{list}}

\newcounter{smallroman}

\newcommand{\ben}{\begin{arabicenumerate}}
\newcommand{\een}{\end{arabicenumerate}}

\author{E.L. Lakshtanov \thanks{FCT-posdoc Aveiro University, Department
of Mathematics, Portugal, e-mail : lakshtanov@rambler.ru}}

\title{Semiclassical limit of the scattering cross section as a distribution}
\date{}

\begin{document}
\selectlanguage{english}
 \maketitle
\begin{abstract}

We consider quantum scattering from a compactly supported
potential $q$. The semiclassical limit amounts to letting the
wavenumber $k \rightarrow \infty$ while rescaling the potential as
$k^2 q$ (alternatively, one can scale Planck's constant $\hbar
\searrow 0$). It is well-known that, under appropriate conditions,
 for $\om \in \bbS_{n-1}$ such that there is exactly one outgoing ray
with direction $\om$ (in the sense of geometric optics), the
differential scattering cross section $|f(\om,k)|^{2}$ tends to
the classical differential cross section $|f_{cl}(\om)|^2$ as $k
\uparrow \infty$. It is also clear that the same can not be true
if there is more than one outgoing ray with direction $\om$ or for
\emph{nonregular} directions (including the forward direction
$\theta_0$). However, based on physical intuition, one could
conjecture $|f|^2 \rightarrow |f_{cl}|^2 + \sigma_{cl} \delta_{
\theta_0 }$ where $|f_{cl}|^2$ is the classical cross section and
$\delta_{ \theta_0 }$ is the Dirac measure supported at the
forward direction $\theta_0$. The aim of this paper is to prove
this conjecture.
\end{abstract}


\section{Introduction}

The semiclassical approximation in physics goes back to the work of
Wentzel, Kramers and Brioullin (WKB) on the Schroedinger equation in
1926. A lot of mathematical work has been devoted to  the subject
and semiclassical scattering has grown into a discipline of its own.
We sketch a simple setup  which is central to our approach.   We
study the scattering of a quantum particle in $\bbR^n$ at a
localized potential $q(x)$. The motion of the particle is governed
by the Hamiltonian \beq H=-\Delta+k^2q(x) \eeq acting on wavefunctions
$\Psi(x)$ in $L^2(\bbR^n)$.

One takes the initial momentum of the particle $k$ to infinity
which is compensated by scaling the potential as $k^2 q$, so as to
keep a balance between kinetic and potential energies. The basic
intuition is that in this limit, the scattering problem is well
approximated by a problem of Newtonian mechanics, namely, the
scattering of a classical particle with momentum $1$ at the
potential $q$.  This was first made precise by Vainberg
\cite{Van0} who proved that for certain outgoing directions $\om
\in \bbS_{n-1}$, \beq \label{intro1} f(\om,k) \approx \sum_{j }
f_j(\om)\e^{\i \theta_j k} +o(k^{-1}),\qquad k \searrow \infty
\eeq In this expression, $f(\om,k)$ is the quantum scattering
amplitude at momentum $k$ . The index $j$ labels different
trajectories which yield the same outgoing direction -$\om$- for
the classical scattering problem and the functions $f_j(\om)$ give
the angular density of trajectories around each of those $\om$-
trajectories. This result has been refined by several authors, we
mention \cite{yajima1,yajima2}, \cite{roberttamura1,roberttamura2}
and recently \cite{Alr}. In \cite{roberttamura1}, it was proven
that under some quite general assumptions, the total \emph{scattering cross section} \beq
\label{intro2}\int_{\bbS_{n-1}} \d \mu(\omega) |f(\om,k)|^2
\rightarrow 2 \si_{cl} +o(k^{0})\eeq where \beq  \si_{cl}=
\int_{\bbS_{n-1}} \d \mu(\omega) |f_{cl}(\om)|^2\eeq is the
classical total cross section (eg \cite[XI.2]{reedsimon})\footnote{In this book the classical total cross section is defined as a measure on  $\bbS^{n-1}\backslash \{\theta_0\}$, that is, a sphere of directions without the forward direction. We understand by total cross section the full measure of   $\bbS^{n-1}\backslash \{\theta_0\}$.}.

The surprising factor of $2$ is well known in the physical
literature and we would like to comment on it. In fact, it appears because of a linguistical problem. Indeed, in
\cite{reedsimon} the classical total cross section is defined so as to measure the 'fraction' of the particles that interact with the scatterer. In quantum
theory, the total cross section measures the defect between the field without
scatterer (incident wave) and the field in the presence of the scatterer
(function $\Psi(x)$). And  these two definitions are not the same! In the shadow zone $\Psi(x)$ vanishes for
any fixed $x$ as $k \rightarrow \infty$ and mathematically this
means that the defect between the wave without a scatterer and in the presence
of the scatterer equals $-1$ multiplied  by the incident wave. So the contribution
to the total cross section of the shadow zone equals  the
geometrical cross section of the support of the potential, which  is $\sigma_{cl}$ according to the
classical definition. So, finally, if one defines the
classical total cross section as the defect between densities of the
free flow of particles and the flow of particles in the presence of the
scatterer, then evidently, the total cross section also equals
twice the geometrical cross section. In this article we will use the
classical definition of the classical total cross section as it is in \cite[XI.2]{reedsimon}, namely $\sigma_{cl}$ equals the geometrical cross section of the  support of the potential.

One of the aims of our article is to state and prove this property rigorously, namely: under the same assumptions (assumptions
\ref{ass: 1}, \ref{ass: 2}) as those required  for (\ref{intro2})  the following
property is valid (see lemma \ref{modva})
\begin{equation}\label{intro22}
\lim_{ \delta \rightarrow 0} \lim_{k \rightarrow \infty}
\int_{\|\omega-\theta_0 \|<\delta} |f(\omega,k)|^2 d S
=\sigma_{cl}.
\end{equation}

Our second aim is to join three facts (\ref{intro1}), (\ref{intro2}), (\ref{intro22}) into one statement.
 Recall that the
result \eqref{intro1} is only valid for certain directions $\om
\in \bbS_{n-1}$. The excluded directions are called
\emph{nonregular}, meaning that classical trajectories accumulate
in those directions. The forward direction is always nonregular,
(since all rays tangent to the boundary of the support of $q$ have
forward direction).

If all  directions, other than the forward one, are regular, then
obviously the forward peak has total intensity $\si_{cl}$, and
hence we might write, in the sense of distributions on
$\bbS_{n-1}$, \beq \label{intro3}|f|^2 \rightarrow |f_{cl}|^2
+\si_{cl}\delta_{\theta_0}, \quad k \rightarrow \infty \eeq where
$\delta_{\theta_0}$ is the Dirac distribution, centered at the
forward direction.\footnote{For hard strictly convex obstacles,
the formula \eqref{intro3} was remarked  in  \cite{sph}, relying
on \cite{Petkov}.} In case when there are two or more rays
scattered into the same direction, the limit of  $|f|^2$ does not
exists. But if we consider $|f|^2$ as a measure on $\bbS^{n-1}$
and supposing that phases $\theta_j(\omega)$ in (\ref{intro1}) are
significantly not coincide (see assumption \ref{ass: 3}) then
formula (\ref{intro3}) is also valid due to the property of
quickly oscillating measure to vanish in the limit. Particularly,
it means that under our assumptions, impact of infinitesimally
small neighborhoods of non regular directions into the total cross
section, goes to zero, as $k$ goes to infinity.

We think that \eqref{intro3} is particularly interesting because
it teaches us immediately that, in contrast to the total cross
section, the \emph{transport cross section}
 \beq \int_{\bbS_{n-1}} \d \mu(\omega)
|f(\om,k)|^2  (1-\cos \om ) \eeq is equal to the classical
expression (in the semiclassical limit.
 This is not at all obvious from the physics point
of view, see e.g.\ \cite[III.A]{Drosdoff} where  one erroneously
concludes that the transport cross section is also twice the
classical value.

Our method of proof is based on the \emph{canonical Maslov
operator}, see e.g.\ \cite{Van1}.

In Section \ref{sec: problem}, we state the problem and result
precisely. In Section \ref{sec: proof}, we present the proof.

\section{Problem and result} \label{sec: problem}
Consider a potential $q \in C_c^\infty(\mathbb R^n,\bbR)$ (smooth
functions with compact support). Choose a unit vector $\theta_0$
in $\bbR^n$ which is to be thought of as the direction of incoming
particles. The projection of $x \in \bbR^n$ on this vector is
denoted $x^n =<x,\theta_0>$ and we write $r:=\strl x \strl = \sqrt{\sum_{i=1}^n      (x^i)^2      }$.

 Define $\Psi(x,k)$ as the unique function in $ C^\infty(\bbR^n \times \bbR^+,\bbR)$  (see e.g.\ \cite{vantychonovsamarsky},  \cite[Add2., Col.2.1]{shubin}) satisfying

\begin{enumerate} \item{ The equation

\begin{equation}\label{odin}
 [\Delta_x+k^2+k^2q(x)]\Psi(x,k)=0, \quad x \in \mathbb R^n,
\end{equation}
}
\item{ The radiation condition
\begin{equation}\label{dva}
u(x,k):=\Psi(x,k)-e^{ik\projx}=f(\omega,k)
r^{(1-n)/2}e^{ikr}(1+O(r^{-1})), \quad r \rightarrow \infty, \quad
\omega=\frac{x}{r} \in \bbS_{n-1}.
\end{equation}
for some function $f(\omega,k)$. }
 \end{enumerate}

The function $f(\omega,k)$, commonly called the scattering amplitude, is uniquely determined by the potential $q$ and
our results will concern its asymptotics as $k \uparrow \infty$.
To formulate our assumptions, we introduce more notation.

\subsection{Assumptions}\label{sec: assumptions}

 Consider the (classical) Hamiltonian $H(x,p)=|p|^2-q(x), \, (x,p) \in
\bbR^n \times \bbR^n$ and corresponding dynamical system
\begin{equation}\label{chet}
\frac{\d x_s}{\d s}=2p_s, \quad \frac{\d p_s}{\d s}=\nabla q(x); \quad
x_0=(y,-a), \quad p_0=\theta_0,
\end{equation}
with $y \in \mathbb R^{n-1}$ and $a \in \bbR^+$ such that $ \sup_{x \in \supp q}( a+\projx) > 0$ where $\supp q \subset \bbR^n$ denotes the support of $q$.
A solution
$t \mapsto (x_t,p_t)= (x_t(x_0,p_0), p_t(x_0,p_0))$  of \eqref{chet} is called a \emph{bicharacteristic}
and its projection to $\bbR_x^n$ (i.e.\ ($x,p) \mapsto x$)  is a
\emph{ray}. Our first assumption expresses that the Hamiltonian
system \eqref{chet} satisfies a \emph{non-trapping} condition,
i.e.\
\begin{assumption} \label{ass: 1} \emph{For any $c<\infty$, there is a $T$ such that
for $s>T$ the rays of (\ref{chet}) with any $y \in \bbR^{n-1}$ are
contained in the region $|x|>c$.}
\end{assumption}
 We denote by $\mathcal I $ the projection of $\supp q$ on the
hyperplane $\projx=-a$.
In accordance with  Assumption \ref{ass: 1}, a ray of \eqref{chet}
with initial data $(y,-a)$ will reduce in finite time to a  line, whose direction is
characterized by the momentum $p_{\infty} ((y,-a),p_0)= \lim_{t\uparrow \infty } p_t((y,-a),p_0) \in \bbS_{n-1}$, since
$|p_{\infty}(y,-a)|= |\theta_0|=1$ by energy conservation.
 This
defines a map \beq J : \mathcal I \mapsto \bbS_{n-1} \subset \mathbb
R^n\, \quad y \mapsto J(y) = p_{\infty}((y,-a),p_0) . \eeq

By $ \left|  \frac{D J(y) }{D y} \right|$, we denote the absolute
value of  Jacobian determinant of $J$. We call $\om \in
\bbS_{n-1}$ a \emph{regular direction} iff. $J(y)=\om$ implies $|
\frac{D J(y) }{D y} |\neq 0$. Otherwise, we call $\om$ nonregular.

\begin{assumption}\label{ass:
2} \emph{ The set $\{ y \in \caI, |\frac{D J(y)}{Dy}|=0 \} $ has
measure zero and
 \beq  \left.\begin{array}{c} y \textrm{ is in the
interior of } \caI \\ y\in J^{-1}(\theta_0) \end{array} \right\}
\Rightarrow  \left| \frac{D J(y) }{D y} \right| \neq 0 \eeq  }


\end{assumption}

 We denote by $\Lam^n \subset \mathbb
R^{n}_{x} \times \bbR^{n}_{p}$ the Lagrangian manifold constructed as

\beq
\Lam^n= \mathop{\bigcup}\limits_{t \in \bbR, y \in \caI}  (x_t((y,-a), \theta_0) , p_t((y,-a), \theta_0) )
\eeq
As global coordinates
on $\Lam^n$, one can choose $(y,t)$ with $y \in \mathbb R^{n-1}$.
By solving (\ref{chet}), we obtain a function  $S=S(x,p)\in
C^\infty(\Lam^n)$: \beq \label{def: S} S(x,p)=-a+\int_{L}<p,dx>,
\eeq
 where $L$ - is the segment of a unique bicharacteristic in $\La^n$ between the
points $((y,-a),\theta_0)$ and $x,p$ for some $y \in \bbR^{n-1}$.
Consider a regular direction $\om_0$.
We can find points (see lemma \ref{generalized} below or Lemma
\cite[lemma 1]{Van0})  $y_1,\ldots,y_v \in \caI$ with
neighborhoods $\caM_i$ such that $J(y_i)=\om_0$ and $J$ is a
diffeoomorphism  on $\caM_i$. Hence on $J(\caM_i)$, we can define
the following map \beq \label{def F}
 J(\caM_i) \mapsto \bbR: \, \om \mapsto F_{i}(\om):=S(x,p=\omega)-<\omega,x>,
 \eeq
 where $(x,p=\omega)$ is a point on the bicharacteristic starting from
 $\caM_i$ and with $|x| >a$. Indeed, for $|x|
>a$, the expression \ref{def F} is independent of $x$ since
$\nabla_x S(x,p) = \omega$.

The next assumption should ensure there are not ``too much"
interference effects
\begin{assumption}\label{ass: 3} \emph{ For any regular value $\omega_0$, the set of critical values of the functions $F_i-F_j$  on $
J(\caM_i) \cap  J(\caM_j)$ has measure zero.}
\end{assumption}

\subsection{Result}\label{res}

The classical differential cross section of the dynamical system
\eqref{chet}, which we denote by $|f_{cl}|^2$,  can be defined as
a distribution on $\bbS_{n-1}$ by the formula
\begin{equation} \label{def fcl}
\int_{\bbS_{n-1}} \varphi(\omega)|f_{cl}|^2 (\omega) \d \mu(\omega) =
\int_{\mathcal I} \varphi(J(y)) dy, \quad \varphi \in
C^\infty(\bbS_{n-1}). \end{equation}
 Note that we denote the Lesbegue
measure on $\bbS_{n-1}$ by $\d \mu(\cdot)$. By Assumption \ref{ass: 2},
$|f_{cl}|^2$ is actually a regular distribution (hence a function), which is known explicitly, see below in \ref{classical f sum of squares}.

We will also need the classical total cross section
\begin{equation}\label{defCTCS}
\sigma_{cl}:= \int_{\bbS_{n-1}} \d \mu(\om) |f_{cl}|^2(\om) .
\end{equation}
From \eqref{def fcl}, it follows that $\sigma_{cl}=meas(\mathcal
I)$ (the Lesbegue measure of $\mathcal I$ in $\bbR^{n-1}$).
 Our result reads
\begin{theorem}\label{thm: conv in distribution}
Let the potential  $q(x)$ satisfy Assumptions \ref{ass: 1},
\ref{ass: 2} and \ref{ass: 3}. Then we have, for all $ \varphi \in
C(\bbS_{n-1})$,
\begin{equation}\label{meq}
\int_{\bbS_{n-1}} \d \mu(\om) \strl f(k,\om)\strl^2 \varphi(\om) \rightarrow
 \int_{\bbS_{n-1}} \d \mu(\om) |f_{cl}|^2(\om) \varphi(\om) +\sigma_{cl} \varphi(\theta_0), \quad k \rightarrow
\infty,
\end{equation}
\end{theorem}

An announcement of this result was published in \cite{lrus}.

 {\bf
Acknowledgments.} The author is grateful to prof. Robert Minlos
for help in preparation of the article.

\section{Proof}\label{sec: proof}

\subsection{Proof of Theorem \ref{thm: conv in distribution}}
\label{sec: proof of main}

The proof of Theorem \ref{thm: conv in distribution} goes through
two lemmas, whose proofs are postponed to the next sections.
\begin{lemma}\label{modin}Assume Assumptions \ref{ass: 1} and \ref{ass: 2}, then
\begin{equation}\label{od}
\sigma=2\sigma_{cl}+o(k^0), \quad k \rightarrow \infty.
\end{equation}
\end{lemma}
\begin{lemma}\label{modva}
Assume Assumptions \ref{ass: 1} and \ref{ass: 2}, then
\begin{equation}
\lim_{ \delta \rightarrow 0} \lim_{k \rightarrow \infty}
\int_{\|\omega-\theta_0 \|<\delta} |f(\omega,k)|^2 d S
=\sigma_{cl}.
\end{equation}
\end{lemma}

For a regular direction $\om$, we introduce an index $j$ which
labels the  elements in $J^{-1}(\om)$. Remark that $J^{-1}(\om)$
is a finite set for regular directions $\om$ since $\caI$ is compact and
$J$ is continuous and therefor pre-images can not be concentred
near the caustic sets. However, the cardinality of $J^{-1}(\om)$
can change.  We put
\beq
\label{classical f sum of squares} | f_{cl}|^2 (\om):= \sum_{j \in   J^{-1}(\om) }
 | f_{j} (\om)|^2, \qquad   f_j (\om): = \left| \frac{D J (y_{j}) }  { D y_{j} } \right| .  \eeq

Pick a test function $\phi$ on $\bbS_{n-1}$ and choose $\ep >0$. Let
$U_1(\ep), U_2(\ep), U_3(\ep)$ be neighborhoods of respectively 1)
$\theta_0$, 2) the nonreguler directions with $\theta_0$ excluded
and 3) the regular directions $\om$ which are critical points of
$F_{j}-F_{j'}$  on $J(\caM_j) \cap J(\caM_{j'})  $(see Assumption \ref{ass: 3})\footnote{ Note that   the index range $j$ and the sets $\caM_j$ in general depend on $\om$.
However, locally the functions $F_j$ are well-defined.  By Assumption \ref{ass: 3}, the set of $\om$ which are critical points of  some $F_{j}-F_{j'}$, has measure zero as a countable union of sets of measure zero.   }.

Choose
the neighborhoods such that $meas(U_{1,2,3}(\ep)) \leq \ep$. We
have to prove that \begin{eqnarray} \label{eq: proof of general
theorem}&& \int \d \mu(\omega) \phi(\om) |f(\om,k) |^{2}  \\&=&
 \int_{ \mathop{\cup}\limits_{z=1}^3 U_z(\eps)} \d \mu(\omega)
\phi(\om) |f(\om,k) |^{2} + \int_{\bbS_{n-1} \setminus
\mathop{\cup}\limits_{z=1}^3 U_z(\eps)} \d \mu(\omega) \phi(\om) |f(\om,k)
|^{2} \\& =& \int \d \mu(\omega) \phi(\om) |f_{cl} |^{2} (\om)+ \si_{cl}
\phi(\theta_0)+ o(1), \quad \eps \rightarrow 0 \end{eqnarray} By
Theorem 2, the fact that $F_{i(\om)}-F_{i'(\om)}$ has no critical
points and \eqref{classical f sum of squares}, the pointwise limit
of the integrand $|f(\om,k)|^2$ in the second term in \eqref{eq:
proof of general theorem} gives $|f_{cl}|^2(\om)$.

Combining now Lemma's \ref{modin} and \ref{modva}, one ends the
proof.
%


\subsection{Preliminaries} \label{sec: preliminaries}

 For any $\omega \in \bbS_{n-1}$, we have the representation (see \cite{Van0}):
\begin{equation}\label{comampl}
f(\omega,k)=\gamma_n \int_{R \bbS_{n-1}} \left [ \frac{\partial
u}{\partial r} + \i k \left < \omega, \frac{x}{r} \right > u
\right ] e^{-ik<\omega.x>} \d \mu(x),
\end{equation}
where $u=u(x,k)$ was defined in \eqref{},
 $R \bbS_{n-1} \subset \mathbb R^n$ is the sphere of radius $R$ and \beq
\label{def: beta} \gamma_n=\gamma_n(k)=-\frac{1}{4\pi} \left (
\frac{k}{2\pi i}\right )^{(n-3)/2}. \eeq Recall the optical theorem
(which could be easy derived using Green formula from
\eqref{comampl}):
\begin{equation}\label{opt}
 Im f(\theta_0,k) = -\gamma_n k \sigma, \quad \forall \, k \geq 0.
\end{equation}

We will need the \emph{canonical Maslow operator}, acting from
$C^\infty(\Lambda^n)$ to $C^\infty(\mathbb R^n)$. We follow the
conventions introduced in \cite{Van0}.

\subsubsection{ The canonical Maslow operator} \label{sec: maslow}
 If the
manifold $\Lambda^n$ can be equipped with the chart ${\mathbb
R^n}_x$, i.e.\  if $x \mapsto (x,p=p(x))$ is a diffeomorphism from
$\mathbb R^n$ to $\Lam^n$, then we can define the canonical Maslow
operator $K_{\Lam^n} : C^\infty(\Lam^n) \rightarrow C^\infty(\mathbb
R^n)$ as
\[
K_{\Lam^n}[\varphi]=I^{-1/2} \varphi \exp(ikS)|_{p=p(x)}, \quad
I=\frac{1}{2} \left | \frac{D(x)}{D(y,s)} \right| , \quad \varphi
\in C^\infty(\Lambda^n),
\]
where $(y,s)$  are global coordinates on $\Lam^n$, introduced in
Section \ref{sec: assumptions}.

It is not always possible to choose $\bbR^n_x$ as global
coordinates since rays can cross.  However, we can fix a locally
finite covering $(\Omega_j)$ of $\Lam^n$ such that for each $j$,
the manifold $\Omega_j$ projects homeomorphically on the subset of
cartesian product of a $l$-dimensional subspace of $\bbR^n_x$ and
a $(n-l)$-dimensional subspace of $\bbR^n_p$. The coordinates in
the respective spaces are denoted as $x_{\al_1},\ldots,x_{\al_l}$
and $p_{\be_1},\ldots,x_{\be_{n-l}}$, ($l=l_j$ here). Hence the
coordinates in the chart corresponding to $\Om_j$ are
$(x_\al,p_\be)$ and the function
\[
I_j=\frac{1}{2} \left | \frac{D(x_\alpha,p_\beta)}{D(y,s)} \right|
,
\]
is bounded away from $0$. Of course, the functions
$x_\beta=x_\beta(x_\alpha,p_\beta),
p_\alpha=p_\alpha(x_\alpha,p_\beta)$ can still be defined.

Let  $\{e_j\}$ be a resolution of  unity on  $\Lam^n$ such that $e_j
\in C^\infty_0(\Omega_j)$ , and let  $\{g_j \in C^\infty({\mathbb
R^n}_x) \}$ be such that  $g_j(x)=1$ in a neighborhood of
${\Omega}_{j,x}$, the projection of $\Omega_j$ on ${\mathbb R^n}_x$,
and such that each $x \in {\mathbb R^n}_x$ belongs to $\supp g_j $
for at most a finite number of $j$.

 The points in $ \Lam^n$ for
which no neighbourhood projects diffeomorphically on ${\mathbb
R^n}_x$, are called \emph{singular}. The projection on $\bbR_x^n$
of the singular points is called the \emph{caustic set}. 

We now define  for each $j$  operator $K_{\Omega_j} :
C^\infty(\Omega_j) \rightarrow C^\infty({\Omega}_{j,x})$ as
\begin{equation}\label{maslov}
K_{\Omega_j}[\varphi]=\left (\frac{k}{-2\pi i} \right
)^\frac{|\beta|}{2} g_j \int_{\overline{\Omega}_{j,p_{\be}}}e_j
e^{ik[G_j(p_\beta)]-i\frac{\pi}{2}\nu_j} I_j^{-1/2} \varphi
dp_\beta,
\end{equation}
where the function $G_j(p_\beta)$ is defined by \beq G_j(x,p_\be)=
S(x(x_\al,p_\be),p
(x_\al,p_\be))-<x_\beta(x_\alpha,p_\beta),p_\beta>+<x_\beta,p_\beta>,\eeq
${\Omega}_{j,p_{\be}}$ is the projection of $\Om_j$ on
$\bbR^{|\beta|}_p$ and $\nu_j$ are the Morse-Maslow-Keller indices
(see details in \cite{Van0}). In \cite{Van0}, it is shown that there
exists a sequence $\eta_{j,m} \in C^\infty(\Lam^n),
m=0,\ldots,\infty, \eta_{j,0} \equiv 1$ such that
\begin{equation}\label{estone}
\Psi_N(k,x)= \sum_{j} \Psi_{N,j}  \qquad \Psi_{N,j}= K_{\Omega_j}
\left [ \sum_{m=0}^N (ik)^{-m} \eta_{j,m} \right ].
\end{equation}
is an approximative solution of (\ref{odin}):  for a compact $V
\subset \mathbb R^n$, one has
\begin{equation}\label{esttwo}
|D_x^\nu[\Psi(k,x)-\Psi_N(k,x)]|<C(V,N,\nu)k^{-N-1+|\nu|+n/2},
\quad k>1,
\end{equation}
If  $V$ does not contain points of the caustic set, than $n/2$ can
be omitted in the RHS of \eqref{esttwo}.

Using \eqref{esttwo}, one can prove the following  celebrated
Theorem
\begin{theorem}\label{thm: vainberg} [Vainberg]
Let $q(x)$ satisfy Assumption \ref{ass: 1} and let $\omega_0$ be a
regular direction, then, for $\omega$ a certain neighborhood of
$\omega_0$
\begin{equation}\label{sss}
f(\omega,k)=\sum_{j \in J^{-1}(\om)} \left|\frac{ D J(y_{j})}{D
y_{j} }\right|^{-1/2} e^{ik(F_{j}(\omega)-\frac{\pi}{2}
\nu_{j})}+O(k^{-1}),
\end{equation}
where $\nu_j$ are the Morse-Maslow-Keller indices  and  the points
$y_{j}$ make up $J^{-1}(\om)$. The functions $J,F$ were
defined earlier.
\end{theorem}
The following lemma is an evident generalization of Theorem \ref{thm: vainberg},
which will be used  in our proofs.
\begin{lemma}\label{generalized}
 Let $y_1,\ldots,y_v$ be points in the interior of $\caI$,
such that $\left|\frac{ D J(y_{j})}{D y_{j} }\right|
 \neq 0$ and $J(y_j)=\omega_0$ for some $\omega_0 \in
S_{n-1} $ (possibly nonregular). Then
 \ben
 \item{
There are neighborhoods  $\mathcal M_j$ of $y_j$ and $R >0$ such
that $J$ is a diffeomorphism from $\mathcal M_j$ to $J(\mathcal
M_j)$ and such that on the bicharacteristics starting from $\caM_j$
and with $|x| >R$, the functions  \beq \mathcal M_j
\times \bbR^+  \to \bbR^+ :\,  (y_j,s) \mapsto \left|\frac{D(x)
}{D(y_j,s)}\right| \eeq exist and are bounded away from zero.  }

 \item{ Define
 \begin{equation}\label{comampl}
f_j(\omega,k):=\gamma_n \int_{R \bbS_{n-1}} \left [ \frac{\partial
\Psi_j}{\partial r} + \i k \left < \omega, \frac{x}{r} \right >
\Psi_j \right ] e^{-ik<\omega.x>} \d \mu(x), \quad
\Psi_j(x)=\sum_{n\geq 0} \Psi_{n,j}(x).
\end{equation}
 For $\omega$ in a certain neighborhood of $\omega_0$
\begin{equation}\label{eq: general vainberg}
f_j(\omega,k)= \left|\frac{ D J(y_{j})}{D y_{j} }\right|
^{-1/2} e^{ik F_{j(\omega)}(\omega)-i\frac{\pi}{2}
\nu_{j}}+O(k^{-1}),
\end{equation}
 }
 \een
\end{lemma}

Statement (1) is an easy analogue of Lemma \cite[Lemma 1]{Van0}.
The only difference is that, where Vainberg assumes $\omega_0$ to
be regular, we simply cut out some bicharacterstics to make the
direction $\om_0$ regular. Statement (2) follows from (1) in the
same way that Theorem \ref{thm: vainberg} follows from Lemma
\cite[Lemma 1]{Van0}. When $\omega_0$ is a regular direction,
Lemma \ref{generalized} reduces to Theorem \ref{thm: vainberg}. In
Theorem \ref{thm: vainberg} there is however no need of
introducing the functions $f_j$.

\subsection{Proof of Lemma \ref{modin}} \label{sec: proof of modin}
Let $h_R$ be a $C^{\infty}$ function on $\bbR$ with support
contained in the interval $[R,R+1]$ and  $\int_{\mathbb R} h_R=1$.

We will estimate
\begin{equation}\label{ftocal}
f(\theta_0)=\gamma_n \int_{\bbR^n} h_R  \left [ \frac{\partial
u}{\partial r} + \i k \left < \theta_0, \frac{x}{r} \right > u
\right ] e^{-ik<\theta_0,x>} \d x,
\end{equation}
for a certain $R>a$.

 Let $ J^{-1}(\theta_0)= \{y_1,\ldots,y_v \}$
and recall (Assumption \ref{ass: 2}) that $J^{-1}(\theta_0)$ lies
in the interior of $\caI$. We choose the  neighborhoods $\mathcal
M_j \subset \mathcal I, y_j\in \mathcal M_j, j=1,\ldots,v$ and $R$
such as in Lemma \ref{generalized}.
We now fix the covering $\Om_{j \in \bbN}$, as required in Section \ref{sec: maslow}.\\

Let for $j=1,\ldots,v$, $\{\Omega_j\}$ be the parts of $\Lam^n$
defined by $y \in \mathcal M_j, |x|>R$.  Let $\Omega_0''$ be the
part of $\Lam^n$ which contains all bicharacteristics with initial
data outside $\caI$ and let $\Omega_0$ be a neighbourhood of
$\Omega_0''$, such that $\Om_0$ does not contain singular points.
This is possible, since the Jacobian $I(x,p)$ is a smooth function and
it equals one for points from $\Omega_0''$, so there exists a
neighborhood where $I(x,p)$ is not equal to zero. Hence $\beta=0$
for $j=0,1,\ldots,v$. The rest of the covering is chosen
arbitrarily, but the functions $g_j$ are chosen such that
 for $|x|>R$, the rays starting from $y_j, j\leq v$ are not in $\supp
 g_j, j>v$. This is possible by Lemma \ref{generalized}.

 Set $\Om_0':= \Om_0 \setminus
\Om_0''$ and let $\widehat{\Om}_0', \widehat{\Om}_0''$ be the
projections on $\bbR_x^n$ of resp.\ $\Om_0',\Om_0''$.

Using $\Psi=\sum_j \Psi_j$, we split the integral (\ref{ftocal})
(changing $\left < \theta_0, \frac{x}{r} \right
>$ into $\frac{\projx}{r}$)

\begin{eqnarray}\label{ftocalDva1}
&&\gamma_n^{-1}f(\theta_0) \\
 &=&  \left( \sum_{j \geq 0} \int_{
\bbR^n } h_R \right) \left [ \frac{\partial \Psi_j}{\partial r} + \i
k \frac{\projx}{r} \Psi_j \right ] e^{-ik\projx} \d x   \\& -&
\int_{\bbR^n}h_R \left [\frac{\partial e^{\i k \projx }}{\partial r} +
\i k \frac{\projx}{r} e^{\i k \projx }\right] e^{-ik\projx}\d x
 \label{long0} \\& =&  \int_{ \widehat{\Om}_0'' } h_R  \left [ \frac{\partial \Psi_0}{\partial r} + \i k
\frac{\projx}{r} \Psi_0 \right ] e^{-ik\projx} \d x  -
\int_{\widehat{\Om}_0''}h_R \left [ \frac{\partial e^{\i k \projx
}}{\partial r} + \i k \frac{\projx}{r} e^{\i k \projx  } \right ]
e^{-ik\projx}\d x   \label{long1} \\
 & +&  \int_{ \widehat{\Om}_0' } h_R   \left [
\frac{\partial \Psi_0}{\partial r} + \i k \frac{\projx}{r} \Psi_0
\right ] e^{-ik\projx} \d x    + \left( \sum_{j >0 } \int_{ \bbR^n }
h_R \right) \left [ \frac{\partial \Psi_j}{\partial r} + \i k
\frac{\projx}{r} \Psi \right ] e^{-ik\projx} \d x \label{long3}
\\
&-&   \int_{\bbR^n \setminus \widehat{\Om}_0''} h_R   \left [
\frac{\partial e^{\i k \projx }}{\partial r} + \i k \frac{\projx}{r} e^{\i
k \projx }\right ] e^{-ik\projx} \d x  \label{long2}
\end{eqnarray}

A first observation is that by application  of  \eqref{esttwo}
with $N=0$, the sum of both expressions  in \eqref{long1} is of
order $-1, k \rightarrow \infty$,  since $\beta=0$ for $\Om_0$ and
$\Psi_{0,0}|_{ \Omega_0 ''} =e^{ik\projx}$ (see beginning of
\ref{sec: maslow})

The term \eqref{long2} is easily seen to give $-2\gamma_n
meas(\mathcal I)k $. Hence, we are left with the two terms of
\eqref{long3}. By using \eqref{esttwo}, these terms are recast in
the form \begin{eqnarray}  \label{eq: proof modin1}  &&k   \int_{
\overline{\Om}_0' } h_R \eta_{0,0} \e^{\i k (S(x)-\projx) }\d x +
O(1) \label{eq: proof modin1} \\ &+ & k \sum_{j>0 }
\sum_{m=0}^{\ell_j} k^{(|\beta|+n-1)/2-m} \int_{ {\Om}_{j,x}
\times {\Omega}_{j,p_{\be}}} h_R \eta_{j,m} \e^{\i k
(G_j(x,p_\beta)-\projx) } \d p_\beta \d  x , \quad k \uparrow
\infty \label{eq: proof modin2}
\end{eqnarray} where $\eta_{j,m} \in C^{\infty}(
{\Om}_{j,x} \times {\Omega}_{j,p_{\be}})$ and the index $\ell_j$
is high enough so as to make the exponent in the error term or
order $O(k^0)$ (since the error term comes from the estimate in
\ref{esttwo} which can be made arbitrarily small by increasing $N$
and hence $\ell_j$.) To show that the term in \eqref{eq:
proof modin1} has order $o(k)$ and the term \eqref{eq: proof modin2} has order $O(k^0)$,
it suffices to note that the critical points of the exponent
$S(x,p)-\projx$ have measure zero, and critical points of the
exponents $G_j(x,p_\beta)-\projx$ have isolated critical points
only. This is shown now.

Using \eqref{def: S}, one calculates \beq   \d (S(x,p=p(x))-\projx
)= <p,\d x>-<\theta,\d x> \eeq which shows that $S(x)-\projx$ has
critical points only at the boundary of $\Omega_0'$, i.e.\ for
$p=\theta_0$. For $j \leq v$, the function $G_j(x,p_\beta)-\projx$
equals $S(x,p)-\projx$ (since $\beta =0$)  and the critical points
$p=\theta_0$ are isolated points in $\Om_j$. The terms in
\eqref{eq: proof modin2} could be calculated explicitly through
the stationary phase method, moreover their leading asymptotics are
given by theorem \ref{thm: vainberg}. Note  that Vainberg
showed in \cite{Van0} that the determinant of the Hessian of
$G_j(x,p_\beta)-\projx$ in the isolated critical points equals  $r^{n-1}\left|\frac{ D J(y_{j})}{D y_{j} }\right|+O(r^{n-2})$
which is not equal to zero according to Assumption \ref{ass: 2}.

For $j>v$  we calculate (for details we refer to \cite{Van0}) \beq
\d (G_j(x,p_\beta)-\projx)= <p,\d x>-<\theta,\d x>+
<x_\be-x_\be(x_\al,p_\be)>
 \eeq
and we find that critical points must again satisfy $p=\theta_0$. By
our choice of the covering $\Om_j$ and the functions $g_j$, $\supp
(h_R \eta_{j,m})$ does not contain points with $p=\theta_0$, since
the projection on $\bbR^n$ of $ \supp \eta_{j,m}$  is contained in
$\supp g_j$. For future use in the proof of Lemma \ref{modva}, we
note that one can continue the expansion up to $\ell_j+1$ to
conclude that for $j>v$, the last term in \eqref{long3} is of order
$O(k^{-\infty})$.

\subsection{Proof of Lemma  \ref{modva}} \label{sec: proof of modva}
Choose the covering
 $\{\Omega_j\}$  and $R>0 $ as defined in the previous
section with the additional constraint about the  $
\widehat{\Om}_0' $ that measure of the set $\{ (y,-a)  \in
\mathcal I : \exists s : (y,s) \in  \projectedmarginalrays\}$ is
smaller than $\delta>0$. This is possible due to the assumption
\ref{ass: 2}.

\begin{eqnarray}\label{}
\gamma_n^{-1}f(\omega) &=& \int_{ \projectedgoodrays }h_R \left [ \frac{\partial
\Psi_0}{\partial r} + \i k \left< \omega, \frac{x}{r} \right>
\Psi_0 \right ] e^{-ik \left< \omega, x \right>} \d x
\label{2long1a} \\
 &-& \int_{\projectedgoodrays } h_R \left [
\frac{\partial e^{\i k \projx }}{\partial r} + \i k \left < \omega,
\frac{x}{r} \right> e^{\i k
\projx }\right ] e^{-ik \left< \omega, x \right>}  \d x \label{2long1b} \\
(\sum_{j>0} f^j:=) \quad  & +& \left( \sum_{j >0} \int_{ \bbR^n} h_R \right) \left [
\frac{\partial \Psi_j}{\partial r} + \i k \left < \omega,
\frac{x}{r} \right>
\Psi_j \right ] e^{-ik \left< \omega, x \right>}  \d x  \label{2long2}  \\
(f^a:=) \quad  &+ & \int_{\projectedmarginalrays} h_R \left [
\frac{\partial \Psi_0 }{\partial r} + \i k \left < \omega,
\frac{x}{r} \right > \Psi_0 \right ] e^{-ik\left < \omega, x \right
>} \d x
\label{2long3} \\
(f^b:=) \quad  &-&  \int_{ \bbR^{n} \setminus  \projectedgoodrays } h_R
\left [ \frac{\partial e^{ik\projx}}{\partial r} + \i k \left < \omega,
\frac{x}{r} \right > e^{ik \projx } \right ] e^{-ik \left< \omega, x
\right>} \d x \label{2long4},
\end{eqnarray}
As in the proof of Lemma \ref{modin}, the sum of terms
\eqref{2long1a} and \eqref{2long1b} is dominated by a constant,
independent of $\om$, and hence these terms vanish upon
integration over a small neighbourhood of $\om= \theta_0$.
Omitting \eqref{2long1a} and \eqref{2long1b}, the above
representation defines the functions
$\{f_j(\omega)\},f^a(\omega),f^b(\omega)$ corresponding to
respectively \eqref{2long2}, \eqref{2long3} and \eqref{2long4}
such that \beq \label{decomposition f} f(\omega)=\sum_{ j
>0} f_j(\omega)+f^a(\omega)+f^b(\omega). \eeq

Call $U(\delta) = \{\, \om \in \bbS_{n-1}, \| \omega-\theta_0 \| \leq
\delta \, \}$. We need to prove that \beq \label{l2 estimate
f2}\lim_{k \uparrow \infty} \int_{U(\delta)} |f(\omega)|^2 \d
S(\om) = \si_{cl}+ o(\delta^0) \eeq Since each $x \in R \bbS_{n-1}
\subset R^{n}_x$ belongs to $\supp g_j$ for a finite number of $j$
and $R \bbS_{n-1}$ is compact, only a finite number of terms are
nonzero in \eqref{decomposition f}. We will show that
\begin{eqnarray}\label{conv of fb} &\lim_{k \uparrow \infty} \int_{U(\delta)}
|f^b(\omega)|^2 \d \mu(\om) = \si_{cl}+ o(1), \quad \delta \rightarrow 0  & \\
\label{conv of fj} & \lim_{k \uparrow \infty} \int_{U(\delta)}
|f^j(\omega)|^2\d \mu(\om) = o(1), \quad \delta \rightarrow 0, \quad
j>0, &
\\ & \label{conv of fa} \lim_{k \uparrow \infty} \int_{U(\delta)}
|f^a(\omega)|^2 \d \mu(\om) = o(1), \quad \delta \rightarrow 0  &
\end{eqnarray} From which \eqref{l2 estimate f2} will follow by
the Cauchy-Schwarz inequality.

\subsubsection{Proof of \eqref{conv of fj}} \label{sec: proof fj}

If $\delta$ is small enough, the set $\Omega_j, \, j>p$ does not
contain bicharacteristics originating from $U(\delta )$. The claim
then follows by the remark at the end of \eqref{sec: proof of
modin}. For $0<j \leq p$, Lemma \ref{generalized} yields
$|f_j(\omega)|< O(k^0)$ .

\subsubsection{Proof of \eqref{conv of fa}} \label{sec: proof fa}

By applying \eqref{esttwo} (with $|\be|=0$)  one obtains \beq f^a
(\om) = \gamma_n(k)\int_{\projectedmarginalrays} h_R(x)
  \widehat{I}(x)   \e^{- \i k <\omega,x>}\d x + o(k^{0} ) \eeq for
$\eta_{0,m} \in C^{\infty}(\Omega_0),\eta_{0,0} \equiv 1$. Here
$\widehat{I}(x)$ is
$$
\widehat{I}(x)= \left ( \frac{d}{dr}+i k <\omega,\frac{x}{r}>
\right )
 \left[ {I}^{-1/2}(x)  \e^{\i k S(x)} \left(\sum_{m=0}^{1}
k^{-m}\eta_{0,m} \right)(x) \right ] =
$$
$$
\e^{\i k S(x)} \left [ ik{I}^{-1/2}(x) \left ( \frac{ dS(x)}{dr} +
\left <\omega,\frac{x}{r} \right >\right ) + \varphi(x) \right]
=\e^{\i k S(x)} \left [ ik{I}^{-1/2}(x) \left ( \left
<p(x),\frac{x}{r} \right >+ \left <\omega,\frac{x}{r} \right
>\right ) + \varphi(x) \right].
$$
Here
$$
\varphi(x)= \frac{d I^{-1/2}}{dr}(x)\left(\sum_{m=0}^{1}
k^{-m}\eta_{0,m} \right)(x) +\frac{1}{k}{I}^{-1/2}(x)
\frac{d\eta_{0,1}}{dr} (x)
$$
finally \beq f^a (\om) = \gamma_n(k)\int_{\projectedmarginalrays}
h_R(x)
  \left [ ik{I}^{-1/2}(x)
\left <p(x)+\omega,\frac{x}{r} \right > + \varphi(x) \right] \e^{
\i k(S(x)- <\omega,x>)}\d x + o(k^{0} ) \eeq

 Now

\beq \label{eq: proof fa} \int_{U(\delta)} |f^a(\om)|^2 \d
\mu(\om) = \int_{\projectedmarginalrays}  h_R(w)
  e^{-\i k S(w)} t(w) \d w  +o(k^{0}) \eeq where for $\quad w \in \projectedmarginalrays$ \beq
\label{def ty} t(w):=(\gamma_n k)^2 \int_{\projectedmarginalrays}
h_R(x) \left [ {I}^{-1/2}(w) \left <p(w)+\omega,\frac{w}{r} \right
> -(i/k) \varphi(w) \right] \eeq \beq \left [ {I}^{-1/2}(x)
<p(x)+\omega,\frac{x}{r}> -(i/k) \varphi(x) \right]
\int_{U(\delta)}   e^{ik[S(x)-<x-w,\omega>]} \d x \d \mu(\om).
\eeq

We can change the order in the  integral due to   integrability of
the density $I^{-1}$ on $\projectedmarginalrays$ (this value is
bounded by $\delta$ due to the choice of $\projectedmarginalrays$
in the beginning of \ref{sec: proof of modva}).

We write $x=r z$ where $r=|x|$ and $z \in \bbS_{n-1}$ and we
perform the integration over the $z$  and $\om$ coordinates. Call
$\tilde{S}=\tilde{S}_{w,r}(z,\om)$ the restriction of
$S(x)-<x-w,\omega>$ to $(z,\om) \in \bbS_{n-1} \times \bbS_{n-1}$.
Since \beq \d (S(x)-<x-w,\omega>) = \langle \d x, p-\om \rangle
-\langle x-w ,\d \mu(\omega)\rangle \eeq one sees that $\d
\tilde{S}$ vanishes whenever  both $p-\om$ and  $x-w$ are
orthogonal to the tangent plane of $\bbS_{n-1}$ in resp.\   $z$
and $\om$, leading to the critical point
   $z^*= z^* (w,r),\, \om^*=\om^*( w,r)$ where
$z^*= \frac{w}{r} $ and $\om^*=p(w)$. By choosing $\delta$ small
enough, this is the unique critical point.

The Hessian matrix of
$\widetilde{S}$ on $\bbR^{n-1} \times \bbR^{n-1}$ is a block
matrix of the form \beq  \mathrm{Hess} \, \widetilde{S}= \left(
\begin{array}{cc}  A & 1 \\ 1 & 0
\end{array}\right) \eeq
where $A$ is the restriction of $\frac{\partial p}{
\partial x}$ to  $| x |=r $ and $1$ is the identity matrix. Since all blocks in
that matrix commute, it is easy to see that this matrix has
determinant $-1$ and hence the unique critical point is nondegenerate. ($\tilde{S}$ is a Morse function)

 Since  $\tilde{S}$ is a Morse function with Hessian uniformly bounded from zero for all values of the parameters $\Sigma, r \in[R,R+1],
w \in \projectedmarginalrays$, there exists  a smooth change of
variables
$(\widetilde{z},\widetilde{\omega})=(\widetilde{z}_{r,w}(z,\omega),\widetilde{\omega}_{r,w}(z,\omega))$
(\cite[v.1 8.2, 8.3, 8.5]{Arnold} {} ))\footnote{see also
\cite[4.8]{Kudr} where are families of functions and germs
deformations are connected} which transforms $\tilde{S}$ to a pure
quadratic form in the neighborhood $\Sigma=\Sigma_{r,w}$ of the
critical point $(z^*,\om^*)$. Since that map is smooth, and $r,w$
vary over a bounded set,  one can bound
\begin{equation}\label{ljacob} \left |
\frac{D(z,\omega)}{D(\widetilde{z}_{r,w}(z,\omega),\widetilde{\omega}_{r,w}(z,\omega)}\right
| \leq C, \quad (z,\om) \in  \Sigma, r \in[R,R+1], w \in
\projectedmarginalrays.
\end{equation}

 Applying stationary phase method  to the expression \eqref{def ty} and using that $w=r z^*(w,r)$, we
get
\begin{equation}\label{estok}
\begin{array}{c}
t(w) = (2\gamma_n k)^2   I^{-1}(w)  h_R(w)
   \\[3mm]
 \int_R^{R+1} \d r   \,  \sqrt{\frac{(2 \pi /k)^{2(n-1)}}{    |  \mathrm{Hess}\,
\tilde{S}_{w,r}(z^*,\om^*)| } }  \e^{\i k ( S(w) +\frac{\pi}{4}
\mathrm{Sgn} H(w,r) )} \left(1+O(\frac{1}{k}) \right)
\end{array}
\end{equation}
where $\mathrm{Sgn}$ stands for ($\sharp$ positive eigenvalues
-$\sharp$ negative eigenvalues). The term $O(1/k)$ in the
last(\ref{estok}) is bounded uniformly on $w \in
\widehat{\Omega}_0 '$ due to (\ref{ljacob}) and absence of the
dependence of the phase on the parameter $r$ and $w$..

Since $(2\gamma_n k)^2 (2\pi /k)^{n-1}=1$  and $ |  \mathrm{Hess}\,
\tilde{S}_{w,r}(z^*,\om^*)| =1$, we have that (\ref{estok}) equals
\begin{equation}\label{estok2}
\int_R^{R+1} \d r h_R(w) I^{-1}(w )
 \e^{\i k (S(w)+ \frac{\pi}{4} Sgn H(w,r))}(1+O(\frac{1}{k}))
 \end{equation}
  Plugging this into \eqref{eq: proof fa} and using $\eta_{0,0} \equiv 1$ (see (\ref{sec:
maslow})), we get
 \beq \int_{U(\delta)} |f^a(\om)|^2 \d \mu(\om) \leq  \int \d r
 \int_{\widehat{\Omega}_0'}(h_R(w))^2 I^{-1}(w)
    \d y  +o(k^{0}) \leq \delta \left (\max_{[R,R+1]}h^2(r) \right ) + o(k^{0}) \eeq

Now, (\ref{conv of fa}) is proved since $\delta$ could be chosen
arbitrary small.

The proof of the statement (\ref{conv of fb}) is a straightforward
application of the stationary phase method.

\end{document}